\documentclass[acmsmall]{acmart}
\usepackage{array}
\usepackage{booktabs}
\usepackage{subcaption}
\usepackage{multirow}
\usepackage{hyperref}
\usepackage{xcolor} 
\colorlet{blue}{black}
\usepackage{booktabs}
\usepackage{tabularx}
\usepackage{ragged2e}
\newcolumntype{Y}{>{\RaggedRight\arraybackslash}X}
\usepackage{longtable}

\AtBeginDocument{%
  }

\acmConference[Conference acronym 'XX]{Make sure to enter the correct
  conference title from your rights confirmation email}{June 03--05,
  2018}{Woodstock, NY}
\acmISBN{978-1-4503-XXXX-X/2018/06}

\newcommand{\parabold}[1]{\vspace*{.5em}\noindent{\textbf{#1.}}}

\begin{document}

\received{May 13 2025}
\received[revised]{January 13 2026}
\received[accepted]{March 17 2026}

\title{Creating Group Rules with AI: Human-AI Collaboration in WhatsApp Moderation}

\author{Gauri Nayak}
\affiliation{%
  \institution{Cornell University}
  \city{Ithaca}
  \state{New York}
  \country{USA}
}
\email{gsn33@cornell.edu}

\author{Farhana Shahid}
\affiliation{%
  \institution{Cornell University}
  \city{Ithaca}
  \state{New York}
  \country{USA}
}
\email{fs468@cornell.edu}

\author{Kiran Garimella}
\affiliation{
  \institution{Rutgers University}
  \city{New Brunswick}
  \state{New Jersey}
  \country{USA}
}
\email{kiran.garimella@rutgers.edu}

\author{Aditya Vashistha}
\affiliation{%
  \institution{Cornell University}
  \city{Ithaca}
  \state{New York}
  \country{USA}
}
\email{adityav@cornell.edu}

\renewcommand{\shortauthors}{Nayak et al.}

\begin{abstract}

WhatsApp is one of the most widely used messaging platforms globally, with billions of users sharing information in private groups. Yet, it offers little infrastructure to support moderation and group governance. In the absence of platform-level oversight, group admins bear the responsibility of governing group behavior. In this paper, we explore how WhatsApp group admins collaborate with AI tools to create, enforce, and maintain group rules. Drawing on a two-phase speculative design study with 20 admins in India, we examine how participants interacted with an AI assistant (Meta AI) to co-create rules and responded to a series of probes illustrating AI-assisted moderation features. Our findings show that while admins appreciated the AI’s ability to surface overlooked rules and reduce their moderation burden, they were highly sensitive to issues of relational trust, data privacy, tone, and social context. We identify how group type and admin style shaped their willingness to delegate authority, and surface the limitations of current chatbot interfaces in supporting collaborative rule-making. We conclude with design implications for building moderation tools that center human judgment, relational nuance, contextual adaptability, and collective governance.

\end{abstract}



\setcopyright{cc}
\setcctype{by-nc-nd}
\acmJournal{PACMHCI}
\acmYear{2026} \acmVolume{10} \acmNumber{6} \acmArticle{CSCW067}
\acmMonth{10} 
\acmDOI{10.1145/3816915}

\maketitle

\section{Introduction}
WhatsApp is the most widely used encrypted messaging platform, where billions of global users share information in unregulated groups~\cite{Morley-2022, Lua-2023}. 
Due to end-to-end encryption, WhatsApp has limited visibility into the content circulating in groups. This places the responsibility of moderation on group admins who are often users who created these groups and informally manage them \cite{shahid2024one}. 

Prior research shows that in formal WhatsApp groups—such as those for work or study—admins tend to actively enforce rules to prevent unwanted content~\cite{Varanasi-2022, shahid2024one}. In contrast, admins of public or close-knit groups (e.g., family and friends) typically avoid moderation, either due to established social hierarchies or the overwhelming volume of problematic content~\cite{Malhotra-2022, Sheryl-2022, shahid2024one}. While the presence of group rules can promote responsible behavior among members~\cite{Dwork-2024, Matias-2019}, most platforms, including WhatsApp, offer little to no infrastructure for rule creation or enforcement~\cite{Wohn-2019, shahid2024one}. Although platforms such as Reddit and Twitch have long used automated moderation tools, they typically rely on human moderators to define rules in advance~\cite{Jhaver-2019}. Even with the growing use of large language models for moderation, the emphasis remains on their ability to detect violations of predefined rules, rather than supporting group admins and moderators in creating or adapting those rules in the first place~\cite{alice-2025}.

Consequently, the process of defining group rules remains largely ad hoc and burdensome \cite{Chandrasekharan-2019, Seering-2019, shahid2024one, Sultana-2022}. Without structural support, admins rely on their own intuition, trial and error, or simply copy rules from other groups \cite{shahid2024one, Wohn-2019}. This lack of support often pushes admins towards reactive approach, where rules are codified only after problematic behaviors
have already damaged the group's social fabric \cite{chadwick2025, shahid2024one}. Given the limited platform support for creating group rules and the reluctance of many WhatsApp admins to engage in proactive moderation, we examine whether AI assistance could help admins establish suitable group rules without adding to administrative workload. 
Specifically, we ask: 

\begin{itemize}
    \item[\textbf{RQ1:}] How do WhatsApp group admins engage with AI to co-create group rules?
    \item[\textbf{RQ2:}] How do admins perceive the benefits and limitations of using AI to create group rules?
\end{itemize}

To answer these questions, we conducted a two-phase speculative design study with 20 WhatsApp admins in India. We focused on India which has the largest WhatsApp userbase with half a billion users~\cite{singh_whatsapp_2025}. In phase 1, participants used the Meta AI\footnote{Meta AI on WhatsApp is an optional assistant developed by Meta that can answer questions, generate ideas, and assist with creative tasks in both individual and group chats. At the time of this study, Meta AI was powered by Meta’s Llama 3.2 model. 
} chatbot within WhatsApp to co-create rules for their groups. In phase 2, they interacted with design probes on different AI-assisted features for creating and enforcing group rules.

Our findings show that the value of AI-generated, data-driven group rules varied across group types, admins' moderation styles, and the relational dynamics within the group. 
Admins of large, formal or semi-formal groups—such as university or neighborhood coordination groups—found AI support helpful in drafting comprehensive rules with minimal effort especially to address issues that often get overlooked due to high message volume. In these groups, where admins had limited personal connection with members, they perceived AI-generated summaries and rule suggestions on potential problematic behavior as effective, amplifying their authority.
In contrast, admins of smaller, informal, and close-knit groups—such as those of friends and family—found little value in AI assistance. These admins relied on relational trust, implicit group rules, and subtle social cues to maintain group order. Thus, they perceived AI-generated rules as unnecessary or even disruptive to interpersonal relationships and group dynamics.

Across diverse groups, admins wanted to engage with AI on their own terms, actively determining which aspects of governance can be delegated to AI and which must remain socially and relationally human. 
They preferred to selectively engage with AI-generated rule suggestions based on their group’s needs, and emphasized adjusting the tone and framing of rules to align with the group’s norms. Many of them also expressed a desire for democratic governance features, pushing against AI systems that only take admin's inputs for rule-making. We discuss how these negotiations are shaped by boundary work around privacy, cultural intelligibility, and accountability in end-to-end encrypted platforms. We translate these insights into design implications for AI tools that enable contextual and collective rule-making to govern encrypted spaces. Overall, our work makes the following contributions to HCI and CSCW scholarship:
\begin{itemize}
  \item Demonstrates the perceived value of AI-assisted moderation in end-to-end encrypted platforms is often contextual and driven by group dynamics;
  \item Surfaces how admins negotiate authority, trust, and relational dynamics when co-creating rules with AI; 
  \item Provides design implications for AI tools that support, rather than supplant, human agency in governing encrypted groups.
  
\end{itemize}

\section{Related Work}
We situate our work within CSCW and HCI literature that examines the social and technical practices of moderation in online communities. We focus on two strands of prior work: (1) how group rules are created, maintained, and negotiated over time; and (2) how AI-driven tools are being introduced to assist moderators.

\subsection{Creating, Maintaining, and Negotiating Group Rules}

Prior work has documented how moderators and group admins---across Reddit, Facebook, Twitch, and WhatsApp---rely on a mix of \textcolor{blue}{formal and informal rules} to guide group behavior. For instance, \citet{Jiang-2020} identified 66 different types of rules in the community guidelines of different social media platforms, addressing online harms, such as revenge porn and graphic violence. \citet{Chandrasekharan-2018} categorized Reddit norms by adoption level, distinguishing platform-wide \textit{macro} rules (e.g., banning hate speech), \textit{meso} rules that are common in large subreddits (e.g., prohibiting spam), and subreddit-specific \textit{micro} norms (e.g., banning Wikipedia links). More recently, \citet{fang2023} extended this work by categorizing rules based on their impact on user behavior, including those that reflect community values (e.g., mutual respect), restrict specific content (e.g., no jokes), and impose structural regulations (e.g., minimum word limits). 
In addition to formal group rules, Reddit and Twitch moderators---as well as WhatsApp group admins---often also enforce implicit \textcolor {blue} {rules} that guide moderation but are not formally codified~\cite{Wohn-2019, chadwick2025}. 

\textbf{Rule creation} is often ad hoc. Although group rules contribute to the stability of online communities~\cite{Dwork-2024, Matias-2019}, most platforms offer little support in creating these rules~\cite{Wohn-2019}. 
In the absence of any platform support and scaffolding, group admins and moderators often draw on their common sense, past experience, and personal values to create rules. Often, they just adopt rules from other groups instead of making their own given the cognitive load involved in creating and enforcing rules~\cite{Chandrasekharan-2019, Seering-2019, Sultana-2022, shahid2024one}. Often, the rules they create are not data-driven and instead consist of generic guidelines discouraging disrespectful attitudes and personal attacks toward other group members.

Group rules are typically made through trial and error~\cite{Wohn-2019}, and new rules often emerge in response to disruptive events, for example, misinformation during the COVID-19 pandemic~\cite{chadwick2025}, or an influx of AI-generated content that admins perceive as low quality~\cite{Lloyd-2025}. \textcolor{blue}{Sometimes, rule design is shaped by technical constraints; for example, Reddit moderators introduce mandatory image captions to support automation via AutoMod~\cite{Jhaver-2019}.
However, in most public and close-knit WhatsApp groups, admins do not make an effort to implement rules against problematic behaviors~\cite{shahid2024one}}. This lack of engagement, combined with limited platform support and the absence of centralized moderation due to encryption, makes WhatsApp a particularly challenging environment for community governance~\cite{udupa2024, shahid2024one}.


Collaborative rule-setting varies across platforms and communities. In some contexts, rule creation involves collective deliberation among moderators—via Slack, Messenger, or Discord~\cite{Wohn-2019, Jhaver-2019, Cai-2022}. In others, especially hierarchical communities, rule-making is more top-down~\cite{Seering-2019}. On platforms like Slashdot, rules are voted on by community members~\cite{Lampe-2004}, but in Reddit and Twitch, involving the wider community is often seen as a risk, especially due to potential rule manipulation~\cite{Juneja-2020, Cullen-2022}.


\textbf{Rule maintenance} is often an ongoing and collaborative process. Reddit moderators constantly update their rules by adding new lists of undesirable phrases or domain names~\cite{Chandrasekharan-2019}. Rule changes happen through reflections on emerging group behavior, especially when new members join the group and act in ways that do not align with the group's implicit \textcolor{blue}{rules}~\cite{Seering-2019, Cullen-2022}. In communities with a clear hierarchy, head moderators often make final decisions and announce rule changes without asking for other moderators' feedback~\cite{Seering-2019}. In contrast, in less hierarchical communities, rule changes typically involve open discussions among moderators, ranging from informal idea sharing to formal debates~\cite{Seering-2019}. Some moderators use community surveys or gather comments from group members on the proposed rule changes~\cite{Seering-2019}. Additionally, each subreddit maintains a Wiki page to keep track of which moderator made what changes to Automod rules so that moderators can be held accountable for their actions~\cite{Jhaver-2019}. 

While most prior work examines how human moderators create, maintain, and negotiate group rules, far less attention has been paid to how these processes might be supported—or reshaped—by AI systems, which we discuss next.



\subsection{AI and Community Moderation}
As moderation becomes increasingly complex, CSCW researchers have begun exploring how AI can reduce the burden on volunteer moderators and scaffold decision-making in online communities. For instance, to help Reddit moderators deal with large volumes of content, \citet{Chandrasekharan-2019} developed an AI tool for flagging content that otherwise went unnoticed by AutoMod. \citet{Schluger-2022} redesigned Wikipedia moderators' dashboard to display which conversations are likely to derail into antisocial behavior. Similarly, \citet{Choi-2023} designed an AI-augmented version of Discord interface that allowed moderators to track activity and toxicity levels for conversations across multiple channels and servers. \citet{koshy2024venire} also developed an AI-guided review panel for Reddit to surface controversial cases so that multiple moderators could review them to reduce inconsistencies in moderation decision. \citet{Khullar-2022} designed AI tools for voice-based discussion forums to relieve moderators of mundane tasks like detecting and rejecting empty recordings. However, most tools assume that group rules already exist, and focus on assisting rule enforcement instead of rule creation. As a result, the use of AI in moderation tasks such as articulating, adapting, or negotiating group \textcolor{blue}{rules}, remains largely unexplored.



Despite this momentum, there is still limited research on how AI can support moderators in the creation of group rules, with one exception being \citet{Xiang-2023}, who designed AI agents to co-create rules in virtual reality communities. Moreover, most of the existing work on community-driven group rules focuses on platforms like Reddit, Twitch, and Facebook. In contrast, very few studies examine how rules emerge or are enforced in end-to-end encrypted platforms like WhatsApp~\cite{norwanto2022, chadwick2025}.


Our work extends this line of inquiry by focusing on how group rules are co-created, adapted, and negotiated in encrypted messaging environments with minimal infrastructure for moderation. While prior research has analyzed how rules emerge through informal practices or are enforced via AI tools, little is known about how admins might collaborate with AI in the actual process of rule creation—especially in high-friction, relationally sensitive settings like WhatsApp which are critically understudied. We contribute to this gap by studying how everyday group admins engage with an AI assistant to draft and modify group rules, surfacing new dynamics of human-AI collaboration in community moderation. In particular, we explore: \textbf{(1) How do WhatsApp group admins engage with AI to co-create group rules, and (2) How do they perceive the benefits and limitations of using AI to create group rules.}
\begin{table}[t]
\centering

\begin{tabular}{@{}m{1cm} m{1.2cm} m{4.2cm} m{2cm} m{2.2cm}@{}}
\toprule
\textbf{ID} & \textbf{Gender} & \textbf{Group Setting} & \textbf{Group Size} & \textbf{Formality} \\
\midrule
P3  & M & University & 50--100 & \multirow{4}{*}{Formal} \\
P11 & M & University & 50--100 & \\
P6  & M & University & $>$100 & \\
P17 & F & University & $>$100 & \\
\midrule
P7  & M & Peer-support group for people with albinism & 20--50 & \multirow{5}{*}{Semi-formal} \\
P15 & F & Former colleagues & 50--100 & \\
P4  & M & Neighborhood residents & $>$100 & \\
P13 & F & CS major students & $>$100 & \\
P19 & F & Group-buy for beauty products & $>$100 & \\
\midrule
P9  & M & Family & $<$20 & \multirow{11}{*}{Informal} \\
P10 & M & Friends & $<$20 & \\
P12 & M & Friends & $<$20 & \\
P18 & F & Friends & $<$20 & \\
P20 & F & Friends & $<$20 & \\
P1  & M & Friends & 20--50 & \\
P2  & M & Job-sharing, Friends& 20--50 & \\
P5  & M & Family & 20--50 & \\
P8  & M & Friends & 20--50 & \\
P14 & F & Friends & 50--100 & \\
P16 & F & Friends & $>$100 & \\
\bottomrule
\end{tabular}
\vspace{0.5em}
\caption{Participant Demographics, grouped by formality and sorted by size}
\label{tab:demographics}
\end{table}

\section{Methodology}

To examine how admins perceive the utility of AI assistance in creating and maintaining group rules, we conducted a two-phase speculative design study with 20 WhatsApp group admins in India. The study protocol received an exemption from the IRB at our institution. 

\parabold{Participant Recruitment} 
We used a combination of convenience and snowball sampling to recruit participants. 
We advertised the study 
on social media groups in India as well as within our immediate network of acquaintances. We selected participants who were at least 18 years old, admins of at least one WhatsApp group, and had experienced problematic content in their groups. 
To ensure diverse perspectives, we recruited admins from both 
public and private WhatsApp groups with varying levels of rule enforcement. These ranged from groups enforcing explicit rules to those where admins relied on implicit \textcolor{blue}{rules}, expecting members to follow them 
without active enforcement. We continued recruiting participants until responses reached theoretical saturation~\cite{pandit-1996}.

\parabold{\textcolor{blue}{Participant Demographics and Group Details}}\textcolor{blue}{Table~\ref{tab:demographics} presents the demographic details of our 20 participants, including seven who identified as female and the rest as male, and aged between 20 to 40 years. All participants owned a smartphone and moderated at least one WhatsApp group. Additionally, all had either completed or were pursuing a college-level degree.}

\textcolor{blue}{Participants in our study were admins of a diverse set of WhatsApp groups that varied in size, purpose, and social formality: formal groups (e.g., university coordination groups), semi-formal groups (e.g., neighborhood associations, peer-support groups, and former colleague networks), and informal groups (e.g., family and friends groups). 
This diversity of group types allowed us to develop a more comprehensive understanding of how AI could support admins across different group types, moderation styles, and social dynamics.
\textcolor{blue}{Participants in our sample also had varying levels of AI literacy, ranging from computer science students and data scientists who were familiar with the use of AI and prompting, to those working in non-tech jobs with limited prior exposure. Despite this variation in AI familiarity, we observed that participants' core attitudes toward the utility and risks of AI-assisted moderation---including concerns about privacy and consent, a desire to retain human authority, and skepticism about AI’s ability to interpret social context---remained consistent across the sample.}
}

\parabold{Speculative Design Study}
We conducted a two-phase speculative design study with 20 WhatsApp group admins to understand what support they might need in creating and deploying rules in their groups. The study was conducted over Zoom either in Hindi or English, depending on the preferences of each participant. After obtaining informed consent, we asked participants in detail about their experiences of handling problematic content in WhatsApp groups,  and their motivations behind setting existing group rules or norms. Then, they engaged in a writing exercise, where they co-created group rules using Meta AI (Phase 1) and shared their reflections on design probes presenting data-driven approaches to rule enforcement (Phase 2). The sessions lasted for approximately one hour and were recorded with the consent of the participants. Each participant was compensated with a gift card worth 500 INR (5.76 USD) for their time and participation. 

\parabold{Phase 1: Co-creating Group Rules with Meta AI} 
To elicit nuanced responses from participants regarding how they perceive the capabilities of AI in helping them make group rules, we conducted a think-aloud writing exercise, where participants used Meta AI to create rules for their groups. We used Meta AI chatbot because it is integrated within WhatsApp~\cite{whatsappfaq} and easily accessible to all our participants. 
In this writing exercise, participants prompted Meta AI to suggest rules for their groups, then, iteratively provided instructions to refine the rules. We asked participants to think aloud as they interacted with Meta AI--describing their expectations, reactions to AI-generated rules, and reasoning behind their modifications to the AI-generated rules. This exercise enabled participants to 
critically engage with the capabilities of existing AI tools when making group rules and reflect on their collaboration with AI in a tangible way. 

\parabold{Phase 2: Design Probes for Rule Creation and Maintenance}
Technology probes are commonly used in HCI to prompt users to critically think about the possibility of new and emerging technologies~\cite{Hutchinson-2003, jarke2018using}. We designed a set of interactive probes to help participants imagine and critique various AI-assisted, data-driven approaches for creating, enforcing, and maintaining group rules---scenarios that would have been difficult to grasp verbally. The probes were modeled after the Meta AI chatbot already integrated within WhatsApp~\cite{whatsappfaq}, ensuring continuity and familiarity, as participants had already interacted with Meta AI in Phase 1 to co-create group rules. 

\begin{figure}
    \centering
     \begin{subfigure}[b]{0.3\textwidth}
         \centering
         \includegraphics[width=\textwidth]{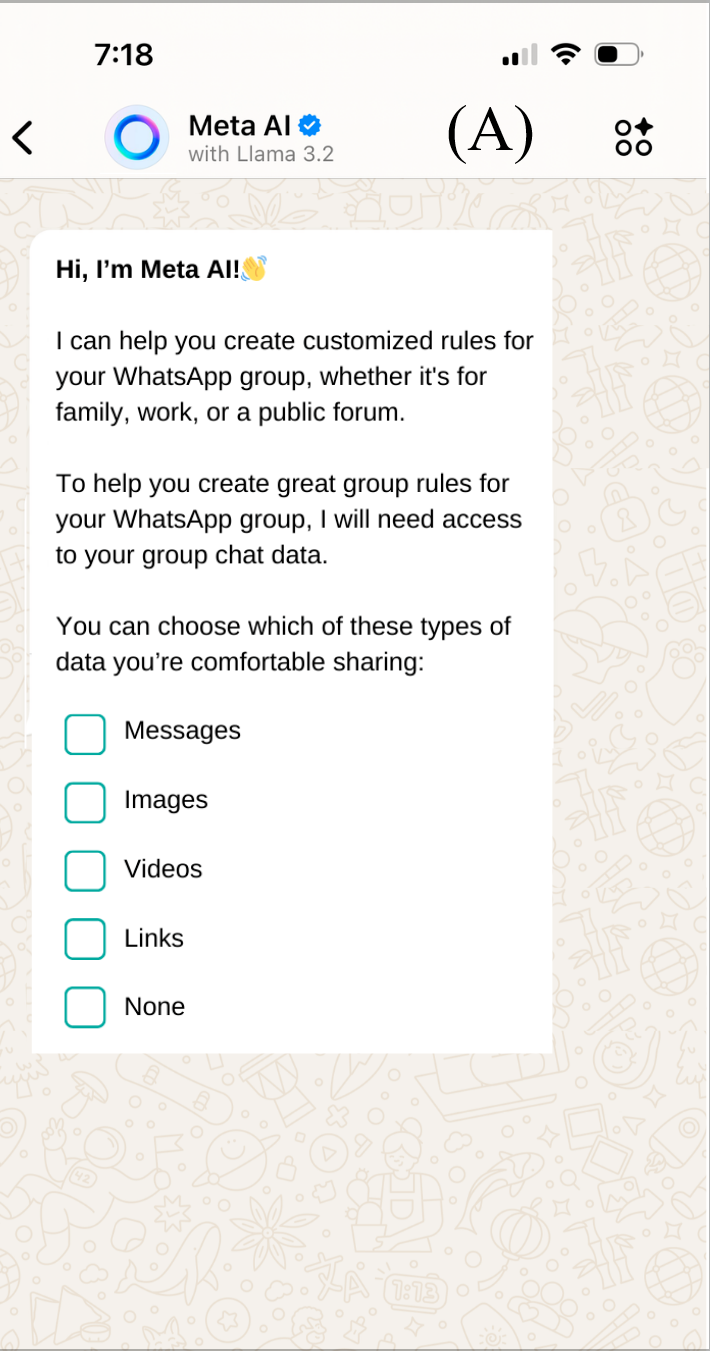}
         \label{fig:figure1}
     \end{subfigure}
     \hfill
     \begin{subfigure}[b]{0.3\textwidth}
         \centering
         \includegraphics[width=\textwidth]{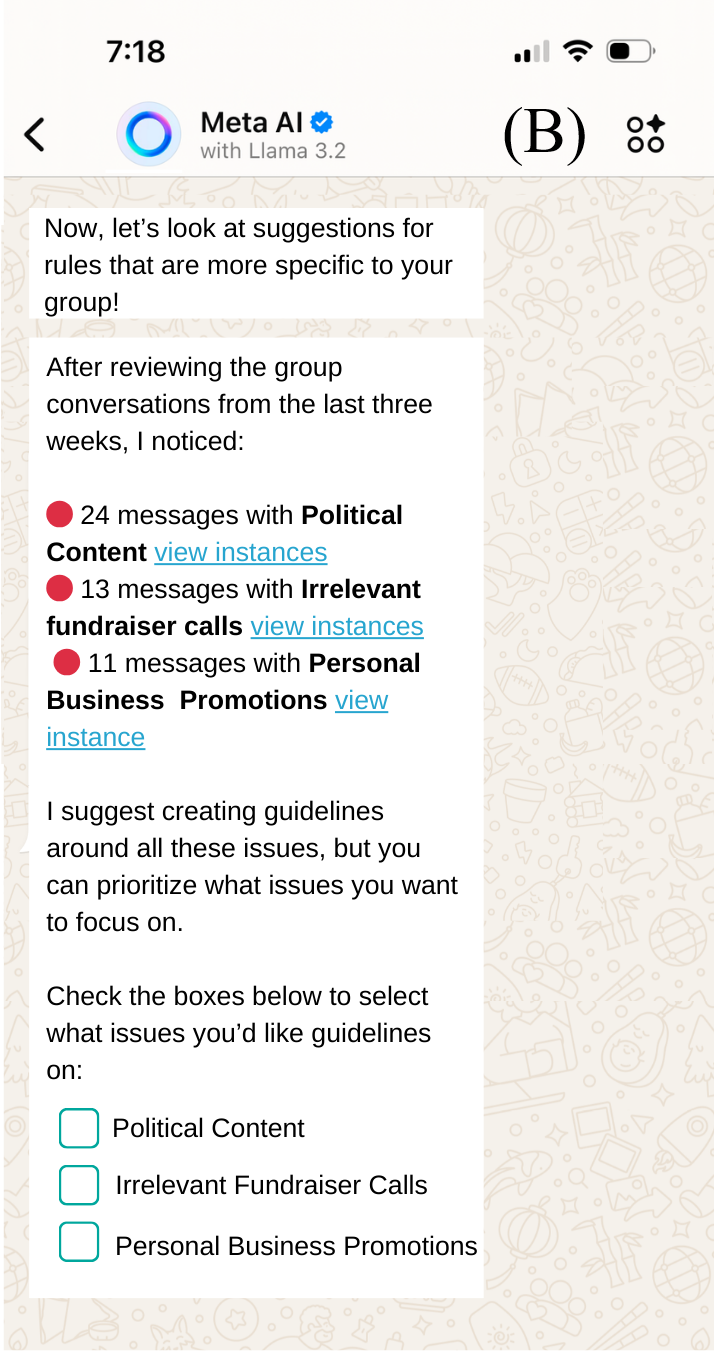}
         \label{fig:figure2}
     \end{subfigure}
     \hfill
     \begin{subfigure}[b]{0.3\textwidth}
         \centering
         \includegraphics[width=\textwidth]{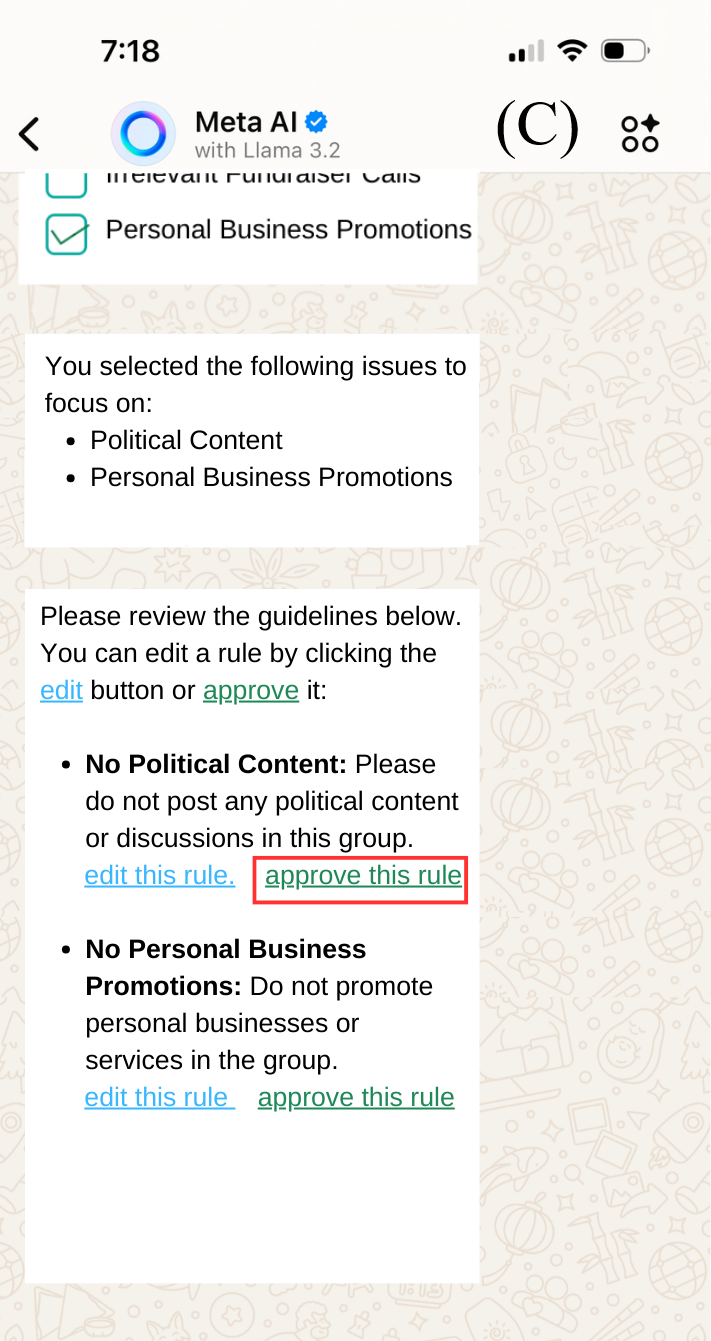}
         \label{fig:figure3}
     \end{subfigure}
          \begin{subfigure}[b]{0.3\textwidth}
         \centering
         \includegraphics[width=\textwidth]{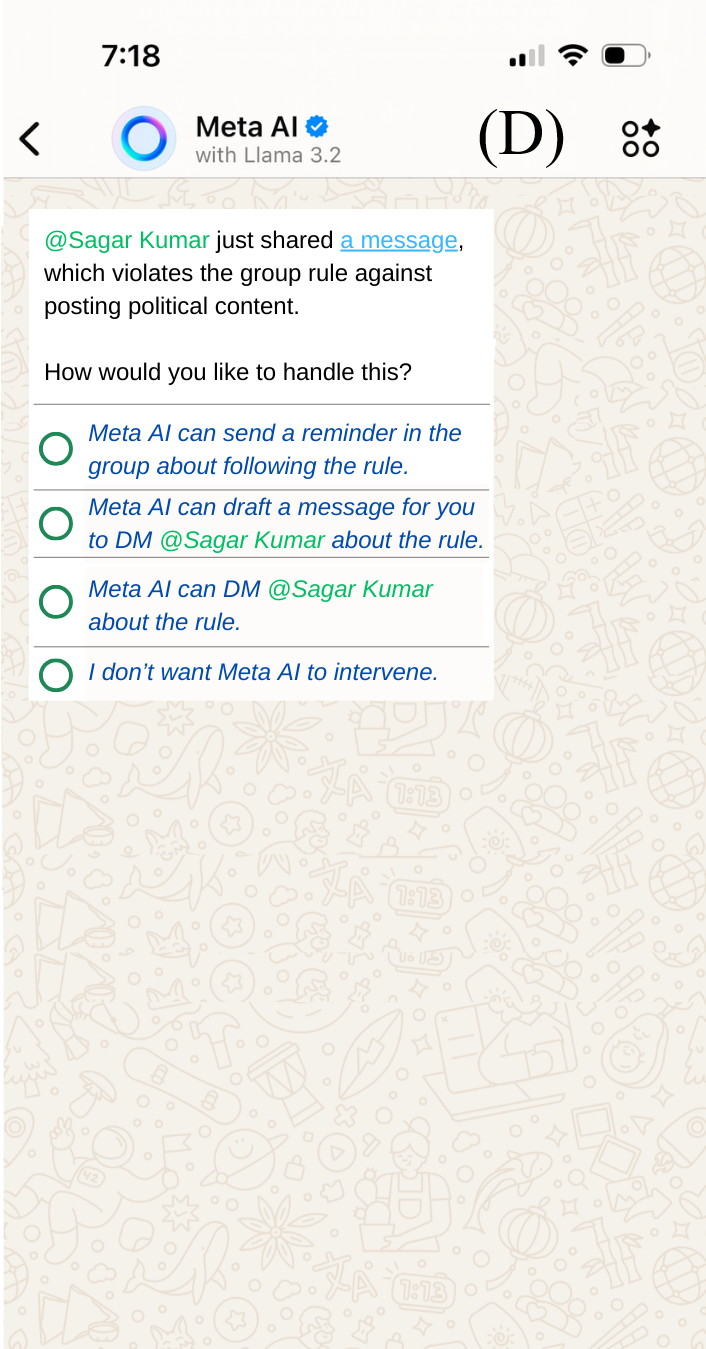}
         \label{fig:figure4}
     \end{subfigure}
     \hfill
     \begin{subfigure}[b]{0.3\textwidth}
         \centering
         \includegraphics[width=\textwidth]{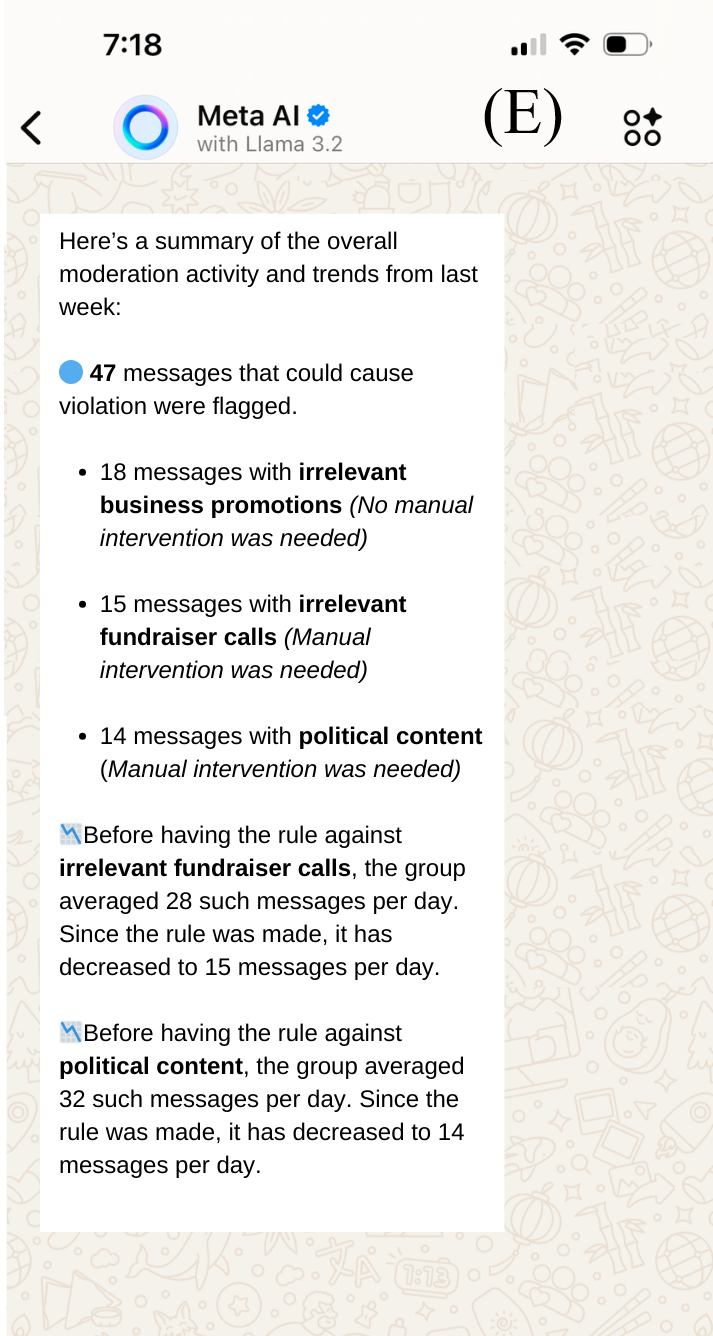}
         \label{fig:figure5}
     \end{subfigure}
     \hfill
     \begin{subfigure}[b]{0.3\textwidth}
         \centering
         \includegraphics[width=\textwidth]{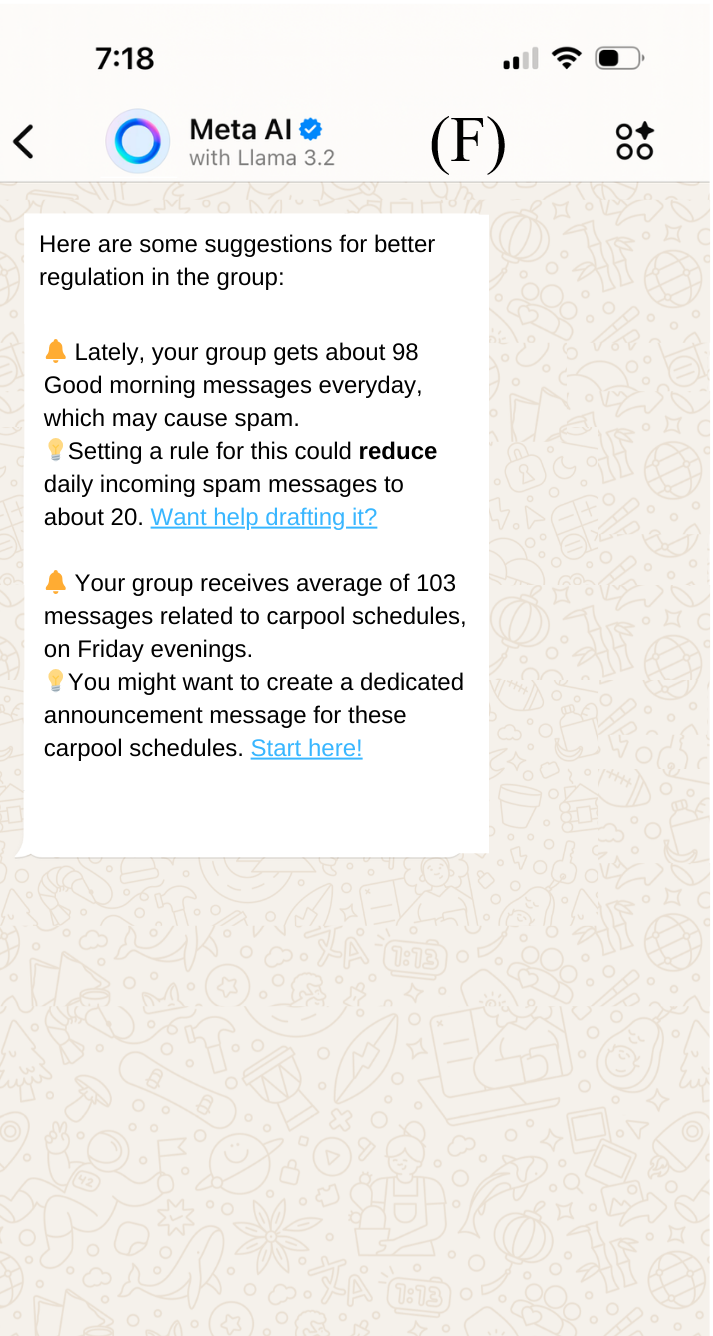}
         \label{fig:figure6}
     \end{subfigure}
    \caption{Design probes illustrating different AI-assisted scenarios for group rule creation, enforcement, and maintenance, used to prompt participants' reflections on AI-mediated moderation.}
    \label{fig:probes}
\end{figure}

The probe \textbf{presented a scenario} in which the Meta AI chatbot asked a WhatsApp admin whether they would like to grant access to the group's content to generate data-driven rules (see Figure~\ref{fig:probes}A). While platforms like Twitch, Reddit, Telegram, and Discord routinely use bots that scan and moderate content in real time~\cite{seering2018social, Kiene2020, alrhmoun2024automating}, WhatsApp lacks such capabilities due to encryption constraints. 
Once granted access, the probe showed the bot generating a brief report on content that might be inappropriate given the group's context (see Figure~\ref{fig:probes}B). 
This scenario 
was informed by prior work~\cite{shahid2024one} showing that WhatsApp admins,  especially in large active groups, often struggle to monitor messages sent to the group. 

Next, the probe showed the bot prompting the admin to review the flagged content and offered suggested rules to address it—rules the admin could accept, edit, or delete (see Figure~\ref{fig:probes}C). 
In a follow-up screen, we explored human-AI interaction for monitoring and enforcing the rules in which the chatbot notified the admin of a rule violation and asked how they wished to respond (see Figure~\ref{fig:probes}D). In this screen, we curated the options based on prior research showing that WhatsApp admins respond to rule violations either publicly or privately depending on their relation with the offender~\cite{shahid2024one}. 

Subsequent screens illustrated additional features: a weekly summary of rule violations and message trends (see Figure~\ref{fig:probes}E), inspired by dashboards used by Facebook and Reddit moderators~\cite{shahid2024one}; and a final screen showing how the AI could detect emerging content trends and proactively update group rules to keep the conversations focused (see Figure~\ref{fig:probes}F).


Participants examined each probe in detail and shared their thoughts on the potential benefits and concerns of using AI-assisted rule enforcement in their groups. At no point did participants share actual group data with us or the probes; they responded to the speculative scenarios and reflected on how they might collaborate with AI to create, adapt, and enforce group rules.

\parabold{Analysis} We collected around 22 hours of audio-recorded data across both phases, along with 132 pages of interaction logs from participants' use of the Meta AI chatbot during Phase 1.  All audio recordings were manually transcribed and translated into English. 

To analyze the interaction logs from Phase 1, we followed Kuckartz's qualitative data analyses method~\cite{kuckartz2019qualitative}. This structured coding process allowed us to explore different stages of rule co-creation process, such as 
providing initial context about the group (e.g., group setting, group's purpose) and customizing the AI-generated rules to fit the group's context. 
Our structural coding involved multiple passes over the data. We applied open coding to develop 14 sub-themes that provided us with a nuanced understanding of participants' interactions with the AI-generated group rules. These include aligning the tone and language to match the group's context, removing irrelevant group rules, among others.

To analyze the interview data collected in Phase 2,  we conducted a reflexive thematic analysis~\cite{braun-2006}.  The first author led the coding process, with all authors meeting periodically to review and refine the codebook, merge overlapping codes, and identify emergent themes. Through multiple rounds of analysis, we developed 107 codes, which were categorized into 12 sub-themes, such as expectations from AI-assisted group rule enforcement and concerns around AI's capabilities. We organized the themes from our data across different stages in the life cycle of group rules, such as creating rules, presenting rules, enforcing and maintaining  rules. \textcolor{blue}{We provide the codebooks used in both phases of analysis in the Appendix.}

\parabold{Positionality} 
Our analysis is shaped by the first-hand experience of inhabiting such WhatsApp groups. Three of the authors are from India and one from Bangladesh. 
All authors have used WhatsApp for over a decade, served as admins of multiple WhatsApp groups, and have firsthand experience with problematic content in these spaces. The authors also have extensive fieldwork experience in India and Bangladesh. Their sociocultural and educational backgrounds align closely with those of the participants, allowing them to understand the underlying dynamics of these groups and elicit nuanced insights.  

We approached this work from an emancipatory action research mindset. We do not assume that AI is a universal solution for moderation, nor do we treat automation as inherently neutral or beneficial. Instead, our goal is to center the perspectives of WhatsApp admins as everyday governance actors, and to critically examine the opportunities, risks, and tensions that arise when AI tools are introduced into these grassroots forms of community management.

\section{Findings}
We begin by describing how participants created, enforced, and maintained group rules, and how these practices varied across different types of groups (Section~\ref{current-practices}). We then present participants’ reactions to the idea of granting AI access to their group messages to generate contextualized rules (Section~\ref{data-sharing}). Finally, we describe how admins envisioned AI support across different stages of the group rule lifecycle, including rule creation (Section~\ref{rule-creation}), enforcement (Section~\ref{rule-enforcement}), and maintenance (Section~\ref{rule-maintenance}).

\subsection{Creating, Enforcing, and Negotiating Rules Across Group Types}
\label{current-practices}
Participants approached rule creation, enforcement, and maintenance in ways that were deeply shaped by the social dynamics and purposes of their WhatsApp groups. Drawing on prior CSCW literature on informal governance and relational moderation~\cite{Seering-2019, Sultana-2022, Wohn-2019, Jhaver-2019}, we identified three broad categories of groups based on the formality of relationships among members: \textbf{formal, semi-formal, and informal}. These categories strongly influenced how admins governed group behavior, whether through explicit rule-setting or tacit social expectations.

\textbf{Formal groups}—such as university coordination groups—tended to have clearly defined, written rules. These groups were usually managed by class representatives and, in some cases, included professors. Rules focused on maintaining order, minimizing distractions, and ensuring productive information exchange. Admins in these groups often created rules collaboratively with co-admins and treated rule presentation as a visible and intentional act: posting rules periodically, pinning them for easy access, and reminding members about them in case of violations.

\textbf{Semi-formal groups} included neighborhood associations, peer-support groups, and professional alumni networks. While members occasionally shared personal updates, the tone of interaction was largely bounded by rules of civility and relevance. Rules in these groups were often written, but their enforcement depended on the admin’s relationship with the group. For example, one admin of a 400-member housing society group in Chennai described enforcing rules that banned political posts and misinformation while still allowing members to share community updates and personal milestones. In another case, a peer-support group for people with albinism used a set of shared \textcolor{blue}{rules} to create a safe and respectful environment for members to discuss both structural issues and personal experiences.

\textbf{Informal groups}, consisting mainly of family and friends, relied heavily on relational trust and implicit norms. 
These groups—typically smaller in size—rarely maintained written rules, even when conversations became heated or off-topic. Instead, admins described using soft interventions, such as private messages or temporary removals, to manage conflicts without damaging interpersonal relationships. In one case, an admin privately messaged a family member who had shared misinformation, gently explaining the issue and requesting they delete the post. Such examples illustrate the emotional and relational labor that admins in informal groups often undertake to preserve group harmony.

While formal and semi-formal groups leaned on structured rule enforcement, including public warnings, deletions, and even temporary group lockdowns (e.g., switching to admin-only messaging), informal group admins avoided confrontational enforcement unless absolutely necessary. Across group types, participants described rules not just as static directives, but as evolving artifacts—shaped through experience, group growth, and emerging issues. For instance, a university admin introduced a rule requiring members to compile technical questions into PDFs after messages became hard to track; another admin prohibited direct messaging between members to prevent harassment, shifting all replies to the public group space.

Admins also differed in how frequently they \textbf{updated} rules. Active admins in formal and semi-formal groups described iteratively refining their rule sets as new challenges emerged. In contrast, passive admins—especially in larger semi-formal or informal groups—typically added or adjusted rules only in response to major disruptions. In several cases, rules were reactive rather than proactive, created in direct response to specific incidents such as harassment or privacy breaches.
For instance, P19, the admin of a group-buy group shared,

\begin{quote}
\textit{I had casually added a non-Indian number, and they sent a long message \ldots{} it seemed like a bot message and included a link. People in the group were worried someone might access their order data and asked me to check who it was, since I don’t regularly check the general chat. Of course, I removed the person from the group and immediately deleted their message. After that, I made a rule that non-Indian numbers can’t join.}
\end{quote}

These findings reinforce prior CSCW work on the socially embedded nature of moderation~\cite{shahid2024one, Seering-2019}, showing how WhatsApp admins draw on interpersonal dynamics, contextual knowledge, and prior experience to shape group governance. They also surface the varying degrees of visibility, formality, and labor involved in creating and maintaining rules, underscoring how moderation on WhatsApp is less about top-down control and more about ongoing, situated negotiation. 

Given this diversity in moderation practices, we next examine how participants responded to the idea of introducing AI into these settings, specifically, their reactions to giving an AI assistant access to group content in order to generate contextualized, data-driven rules.

\subsection{Negotiating Data Access for AI-assisted Moderation}
\label{data-sharing}
In Phase 2, we explored participants' attitudes toward granting the Meta AI chatbot access to their WhatsApp group content---such as messages, images, and shared links---in order to generate contextualized, data-driven rules (Figure~\ref{fig:probes}A). Participants’ responses revealed a complex set of sociotechnical considerations grounded in group context, perceived privacy risks, and collective responsibility. 

Admins of large, formal groups—such as those used for academic coordination or workplace logistics— 
were open to providing the AI with conditional access to group data, but only if key privacy concerns were addressed. These groups typically focused on less sensitive topics, such as academic queries, assignments, exam schedules, or organizational announcements, which admins felt posed minimal privacy risk. However, despite their openness, many admins felt they did not have the right to grant AI access to group data on their own and emphasized the importance of member consent and collective data governance.  Several participants felt uncomfortable making unilateral decisions on behalf of their groups, and instead believed that any AI access to group data should be subject to explicit, revocable consent from members. 
Participants also described situations where access should be time-bound or reversible, such as during politically sensitive periods (e.g., elections) or emotionally charged discussions. These concerns reflect how admins understood group privacy as context-dependent, shaped by both the nature of the group and the evolving sensitivities of its members.

Some participants expressed willingness to share group data if the AI's data usage policies are understandable and transparent. P4, the admin of a semi-formal, large neighborhood group echoed broader concerns around data locality and reducing reliance on opaque cloud-based systems:
\begin{quote}
    \textit{I would be less concerned about the data privacy if the entire processing is happening on my device and the data doesn’t go out.}
\end{quote}

Across group types, admins also \textcolor{blue}{anticipated distinguishing} between types of data they would or would not share with AI based on perceived privacy risk. Almost all participants strongly opposed sharing sensitive user-generated content such as messages about political views, health information, or private data like home addresses. In contrast, there was broad agreement that external links to news articles, YouTube videos, or social media posts could be shared without concern, as they did not reveal personal data. In these cases, admins wanted the AI to provide additional support when processing these links, such as flagging external links to malicious content, summarizing news articles behind the paywall to improve accessibility, and fact-checking potential fake news. Participants also considered group metadata, such as group's size, purpose, or descriptions as low-risk and sufficient for AI to provide relevant rules, without requiring additional access to group messages. As P2, the admin of a large group of friends, explained: 
\begin{quote}
\textit{I am okay with sharing some group info because we already have those in the group description and group name. So maybe AI can read these and have some background data[...] I guess I could give access to messages and links, but being an admin, I also have to think about group members' private images and videos. So I may not agree with sharing that content with AI.}
\end{quote}

These findings show that participants' willingness to grant AI access to group data is shaped by their attitude towards member consent, perceived risk of sharing different types of data, and  how those data will be processed. This careful boundary work reflects the nuanced sociotechnical roles that admins play as informal data stewards. 

Next, we describe how participants envisioned the role of AI across the different stages in the group rule lifecycle—creation, enforcement, and maintenance.

\subsection {AI and Rule Creation}
\label{rule-creation}
In Phase 1, participants interacted with the Meta AI chatbot to generate rules for their existing WhatsApp groups. This process surfaced how admins from different group types engaged with AI-generated suggestions, negotiated tone and relevance, and wrestled with the limitations of current chatbot interfaces. Participants noted both the advantages and frictions of using AI to draft group rules—highlighting where the tool aligned with their needs, where it fell short, and how its usefulness varied across group contexts.

\parabold{Expanded Rule Coverage, But Limited Specificity}
\textcolor{blue}{During phase 1,} most participants provided minimal input when prompting Meta AI to create group rules, often only mentioning the group’s purpose without further context about size, dynamics, or moderation preferences. Prompts such as \textit{“Can you help me create rules for my college group?”} or \textit{ “I want some rules for friends' group”} were common starting points. 
The initial AI-generated outputs were typically verbose, used formal language, and contained 12 to 15 rules structured into segments like `General Rules', `Posting Etiquette', `Content Guidelines', `Admin Responsibilities', and `Consequences'. Irrespective of the group type that participants mentioned in their prompts, Meta AI suggested generic rules like “Be respectful and inclusive,” “Members must follow admins’ instructions,” or “Members will be removed after repeated violations.”

Participants appreciated that even with minimal input, the AI could generate not only expected rules but also suggest new ones they hadn’t previously considered. These unexpected suggestions surfaced \textcolor{blue}{rules} that had been overlooked or were difficult to articulate, offering admins a way to reflect on and formalize emerging concerns. For example, P7, the admin of a peer-support group for people with albinism---a condition that affects vision and skin pigmentation---was impressed when the AI suggested a rule related to audio messages: \textit{"Keep audio messages short and clear. Use a slow and clear pace for easier comprehension."} Although this rule was not part of existing group rules, P7 found it highly relevant given that group members frequently used audio messages. Similarly, P14, who managed a formal university group, appreciated a rule about responsible emoji use. While her group had not formalized such a guideline, members had previously misused emojis to insinuate inappropriate meanings. She found the AI’s suggestion timely and planned to incorporate it. She described:

\begin{quote}
    \textit{It would have taken me about an hour or two to completely think of all the hypothetical cases that could happen and their consequences before making these rules. But here it just took me a few seconds to  generate it because the AI was intelligent enough to understand what I'm trying to say.}
\end{quote}
This shows that while active admins could have crafted rules themselves, the AI’s ability to surface relevant rules, including the ones often overlooked, made the process faster and less effortful.

Several participants noted that the AI’s generic outputs often overlooked key contextual rules they had already established. 
For instance, P19, who managed a group to buy and sell beauty products, forgot to include purchasing instructions that she had listed in her existing group rules. Similarly, P2, the admin of an informal friend group meant for job-sharing, failed to include the rule banning NSFW content, which he had introduced in his group after a prior incident. While the AI-generated rules like \textit{“no spam”} or \textit{“stay on topic”} could apply to his needs, these rules lacked the specificity to address the problem the group encountered previously. This suggests the need for scaffolding or explicit nudges from the AI to capture existing group \textcolor{blue}{rules} within suggested rules.

\parabold{Iterative Editing and Tone Adjustment} Admins who were already active moderators—particularly in formal and semi-formal groups—were more likely to engage critically with the AI’s output. They refined rules through iterative prompting, asking the AI to  incorporate existing \textcolor{blue}{rules} or adjust phrasing for clarity and authority. 
For example, P13, the admin of a large computer science student group focused on technical discussions and internship opportunities, asked the AI to include a rule about \textit{“no personal messaging within the group”}--reflecting the group’s existing policy prohibiting unsolicited direct messages to unknown group members. Similarly, P4, an admin of a neighborhood group, revised a blanket ban on business promotions to allow those from within the community. 
These examples show how experienced admins treated the AI as a writing assistant, iteratively shaping its outputs to align with established community values. These contextualized inputs led to more tailored and relevant rules in subsequent iterations.

Admins also refined the tone embedded in the rules. For example, admins of formal groups instructed the AI to adjust the language to be more formal and strict, believing that the neutral, default tone of AI-generated rules would not be taken seriously by group members. For instance, P3, an admin of a university students group remarked that the initial AI-generated rules were \textit{``very lenient.''} He then asked the AI to\textit{``give much stricter rules along with consequences so that members keep them in mind while sending messages.''} 

In contrast, admins of informal, close-knit groups preferred a casual tone and saw the rules as merely suggestive. They often accepted the AI’s initial output with only surface-level changes—such as making the language friendlier or shortening the list. 
Several admins voiced frustration over long, redundant rule lists and asked the AI to \textit{``remove generic rules.''} They requested shorter and more concise rules capturing the essential points so that group members could review them easily. For example, P18, who managed a small informal friends group, asked the AI to give only \textit{``three-four one line rules''} so that she can \textit{``add the rules to the WhatsApp group description.''}

Almost all participants faced difficulty when editing AI-generated output due to its lengthy, unstructured format. When participants wanted to edit a particular rule, they had to scroll through the long output, locate the specific rule, and then refer to the rule using keywords if the outputs were not numbered. Although occasionally having rule numbers made it easier for participants to prompt, \textit{“change rule number X”}---this did not always work when there were multiple sub-rules under a specific rule, making it difficult to indicate which segment needed editing. Therefore, participants expressed the need to be able to select a specific rule for modifying it directly. In cases where the changes were minor, such as deleting a rule entirely or changing a phrase, several participants felt it would be faster and easier to directly edit the output themselves, rather than prompting the AI. Additionally, the back-and-forth interaction with the chatbot made it difficult to compare how the rules changed over multiple iterations. As a result, participants often relied on memory or repeated similar commands multiple times to achieve the desired tone. 

\parabold{Formatting and Presentation of Rules} 
When participants were satisfied with the final set of rules, they tried to make them more presentable by focusing on formatting. They wanted a version that could be easily copy-pasted in the group, without any signs of being generated by AI because the output included statements like “Here are your rules” or “Would you like to modify them?” P8, the admin of a friends group explained: 
\begin{quote}
    \textit{``I am expecting it [Meta AI] to remove the first line and give me the final version. I'm hoping it to be a singular message that I can paste in the group.''} 
\end{quote}

Participants also asked for cleaner layouts---removing unnecessary line breaks, fixing misnumbered items, bolding key points, and deleting excessive spacing. In informal friends groups, admins also wanted the AI to add appropriate emojis to the rules to match the group's \textit{``vibe.''} Some admins even asked for a text-based image of the rules to share easily, due to the character limit in WhatsApp group descriptions. However, the AI sometimes returned irrelevant images (e.g., stock photos of people in meetings) highlighting current limitations in aligning output formats with user intent. These findings highlight the extra \emph{presentation work} the admins needed to do to make rules legible, acceptable, and usable for their groups. 

Admins with limited English proficiency found the AI particularly helpful in articulating rules clearly and concisely. They appreciated the AI’s ability to restructure long sentences and simplify language, making the content accessible for both them and their group members. P19 shared:
\begin{quote}
\textit{``I'm not good at framing my words properly to communicate with people... Meta was helpful in this case.''}
\end{quote}

\parabold{Skepticism about Formalized AI Rules}
Admins of close-knit, family and friends groups often wondered if written rules were necessary to promote group harmony since members already knew each other. Some participants \textcolor{blue}{anticipated} that introducing structured rules might seem \textit{``too formal''} and \textit{``out of place''}, making members question the rules, or even the intimacy of the group itself. 
For instance, in the case of P9, Meta AI recommended a rule for their family WhatsApp group stating \textit{“members should support each other through thick and thin.”} P9 found this rule performative and asked the AI to remove it, noting that this kind of emotional support is implicit among family members and does not need to be formalized through a WhatsApp group rule. Similarly, when P8 asked for rules for their friends group, Meta AI provided a rule outlining when members should be permanently banned from the group. P8 removed this rule, explaining that he never anticipated to take such harsh action, because \textit{"they are all good friends."} He elaborated:
\begin{quote}
\textit{"I'm pretty sure the moment I'm going to post this, I'll be mocked because we don't have any written rules\dots People will ask why are you making this? Even though these rules are quite friendly and not strict, people in my group might mock me."} 
\end{quote}


Overall, participants viewed AI as a helpful starting point for rule creation, especially in surfacing overlooked \textcolor{blue}{rules} and reducing the time and effort needed to draft comprehensive rules. But co-creating rules with AI was not a simple plug-and-play process—it required admins to continuously adapt, edit, and format the output to make it usable and appropriate for their groups. In high-trust groups, formal rules could feel out of place. In larger or more structured groups, rules needed to convey authority and clarity. These findings highlight that rule creation is not just a technical task, but a situated practice shaped by social expectations, aesthetic preferences, and the admin’s role as both facilitator and gatekeeper. In the next section, we examine how participants imagined AI supporting the enforcement and maintenance of rules after they are introduced.


\subsection{AI and Rule Enforcement}
\label{rule-enforcement}
Participants' responses revealed that the perceived utility and risks of AI-mediated enforcement varied across group types, moderation styles, and relational dynamics.

\subsubsection{AI Overview of Problematic Content}
In phase 2, we presented participants with AI-generated reports summarizing potentially problematic content in their groups (Figure~\ref{fig:probes}B). Admins, especially those managing large, active groups, \textcolor{blue}{anticipated value in these reports as a way to monitor alignment with group rules} and assess whether conversations were drifting off-topic. For example, P15, the admin of a semi-formal ex-employees group, said that if the AI flagged frequent political discussions—off-topic in her group—she would use that insight to reinforce or introduce a rule against political content. Similarly, P13, the admin of a large university group, noted that such reports could help surface content like irrelevant promotions or ads that might otherwise go unnoticed due to high message volume. 

Admins also saw value in using AI-generated reports to track the frequency of rule violations and identify new disruptive behaviors that might require intervention.  Some noted that such summaries might reveal areas where new rules were needed or where existing ones were too rigid. P3, the admin of a formal university group, pointed out that if many members were posting a certain type of problematic content that isn't explicitly banned, it could indicate the need to either establish a new rule or create a separate group for those discussions. Similarly, P8, the admin of a friends group, expressed that the AI reports could inform whether the existing group rules are actually relevant or overreaching, depending on how group members were actually using the group. 

At the same time, some admins expressed concern that an AI-generated report that merely listed the instances of problematic content, could be misleading without additional context. For example, if the AI reported that the group received a large amount of problematic content of a certain kind, admins wanted to know whether these messages were all posted on a single day or over a span of days, as this would significantly affect how urgent the situation is. Some admins felt that the AI should also show which group members were sharing problematic messages to help them identify members with disruptive behavior. 

Building on these concerns, many admins pointed out that the appropriate enforcement of rules requires understanding the deeper context of the flagged content. They wanted the AI to do more than just identify the number and categories of problematic content as shown in the probe. Many also wanted the AI to go beyond categorizing content and instead interpret tone, intent, and cultural nuance. For instance, they wanted the AI to distinguish between sarcasm and hostility, or between casual political jokes and propaganda, though they also questioned the AI's ability to accurately assess these nuances, especially when the content was not in English, as was common in some of the groups. P19 shared:
\begin{quote}
\textit{``Sometimes in my group, there are people who joke a lot, use sarcastic comments, and AI can always detect that and may potentially flag it as spam or something like that. Or when people don't use English or use references to movies or memes, it might get flagged as an offensive message or something. That could be an issue.''}
\end{quote}
 
In contrast, admins of smaller, informal groups found these reports less necessary and felt rule enforcement to be manageable.  
As P8, admin of a friends group put it, 
\begin{quote}
\textit{``It is a good-to-have, I'm not really a big fan of these kinds of things \dots{} If the group is large, then maybe these are good enough, but in a group of 15 to 20 college friends this feels a bit extra—as if I'm trying to micromanage the group.''}
\end{quote}

These findings suggest that the value of AI-generated content reports is highly context-dependent. While useful for flagging trends in large, fast-moving groups, such summaries require additional nuance and customization to be effective across diverse settings.

\subsubsection{AI-Assisted Interventions for Rule Violations}
We then explored how participants felt about AI intervening directly after a rule violation, either by sending a group-wide warning or by privately messaging the offender (Figure~\ref{fig:probes}D).

\parabold{Private Messaging}
In close-knit groups, admins preferred to personally reach out to violators rather than delegating the task to AI to preserve interpersonal relationships. Admins in these groups worried that the AI-issued warnings would feel impersonal, lacking the nuance and warmth needed to carefully handle violations caused by members they knew personally. As P10, the admin of a small friends group, explained:
\begin{quote}
\textit{``I would like to have a better human touch to interact with my friends. It's a friends group. If I select the AI to DM my friend, it will be more of an AI censorship.''}
\end{quote}

This sentiment was especially strong when addressing minor or first-time violations. Admins preferred drafting a personalized message with the help of AI over allowing the AI to directly reach out to violators. They also wanted to review and edit any message before sending it, particularly in cases involving friends or family. Admins also preferred sending the warning message themselves via DMs when dealing with repeat offenders, seeing it as a more deliberate escalation before considering harsher actions such as temporary or permanent removal. 

In formal groups, some admins lacked confidence in the AI’s ability to convey the gravity of rule violation to offenders. They worried that the AI would default to a neutral tone and its warnings would be perceived as insincere and weak. They believed personal outreach from the admin would signal more authority and would be taken more seriously by the offender. Moreover, some admins didn't trust the AI's ability to correctly interpret the context of rule violations. Instead, they preferred to directly engage with the violators to better understand their intent before taking any action.

In contrast, admins of large, public, or loosely moderated groups welcomed the idea of AI-led messaging—particularly when they did not know the violator personally or wanted to avoid confrontation. This feature was particularly welcomed in groups where confronting offenders might expose admins to harassment due to gender dynamics or social hierarchy. In these cases, admins saw AI as a protective buffer between themselves and the offenders. For example, P13, a female admin of a large group of computer science students shared:
\begin{quote}
    \textit{``Most members in my group are male students who occasionally post adult humor against the group rule. I could use an AI to message the violators I don't know personally. Because if I message them myself, they might harass me.''}
\end{quote}

In smaller, close-knit groups, a few admins also found it easier to offload rule enforcement to the AI to avoid directly moderating people in their social circles. For example, participants felt confronting a close friend for violating a rule might lead to awkwardness. Similarly, correcting an elderly relative in a family group might make them feel disrespected by the admin. In such situations, admins appreciated that the AI could play the \textit{“bad cop,”} taking the burden of rule enforcement from the admin, while preserving social harmony. 

\parabold{Public Warning}
Some admins managing large, formal groups preferred that the AI send public, group-wide warnings after a violation. They perceived that these reminders will allow them to address violations collectively without singling out individual. They hoped this would signal to members that the group was regularly moderated and any violations would be promptly addressed. They felt such warnings would be more effective especially when multiple members engaged in similar disruptive behavior, reducing admins' workload to reach out to individual violators. 

However, in informal or smaller groups, admins found public warnings unnecessary and even counterproductive. They worried that such messages would feel too formal, disrupt ongoing conversations, or embarrass the violator in front of peers. As one admin noted, a group-wide warning might come across as \textit{``passive-aggressive or controlling in a tight-knit group.''} In these settings, casual, private nudges were preferred. 

These findings illustrate that AI-assisted rule enforcement is not one-size-fits-all. Admins navigated a complex mix of relational dynamics, group size, and social norms when considering how and when to delegate enforcement to AI. While some saw AI as a helpful assistant or buffer, others viewed it as a potential disruption to group culture. Designing AI tools that can adapt to these contextual differences—and support both soft and strict forms of enforcement—is critical to their acceptance and effectiveness.


\subsection{AI and Rule Maintenance}
\label{rule-maintenance}
Beyond creating and enforcing group rules, participants also reflected on how AI could support the ongoing work of maintaining and updating rules over time. This included interpreting whether rules were working as intended, identifying new patterns of behavior that might warrant policy changes, and helping admins adapt rules as group dynamics evolved. We explore two proposed AI features—weekly moderation summaries (Figure~\ref{fig:probes}E) and AI-suggested new rules (Figure~\ref{fig:probes}F)—and how participants engaged with their possibilities and limitations across different types of groups.

\subsubsection{AI-generated Weekly Summary}
In Phase 2, participants were shown probes depicting a weekly AI-generated summary of moderation activity. These included data such as the number of messages flagged for violating rules and trends indicating how group behavior changed following rule implementation. Most admins who managed large, highly active groups, expressed strong interest in using AI-generated moderation summaries. They described feeling overwhelmed by message volume in their groups and saw summaries  as a way to gain a high-level view of how the rules were impacting group behavior, and  identify emerging trends in rule violations.

However, even among those who found the summaries useful in theory, few planned to use them actively. This was especially true for passive admins, who only engaged in moderation in response to disruptive group behavior or when group members brought an issue to their attention. In small, close-knit groups, where violations were rare and moderation was minimal, admins felt such features added little value. They wondered if merely receiving data-driven insights was enough for decision making. 
For example, P11, the admin of a university group remarked:
\begin{quote}
    \textit{``What ultimately matters is whether the chat feels clean, not the raw statistics on how many violations occurred since a rule was implemented.''}
\end{quote}

These observations suggest that apart from focusing on counts or trends, AI systems should consider how to meaningfully convey the impact of rules in ways that resonate with  WhatsApp group admins’ intuitive judgments and decision-making styles.

\subsubsection{AI-generated Suggestions for New Rules}
We also presented AI-generated suggestions for new rules based on recent group activities (Figure~\ref{fig:probes}F). Admins of large, highly active groups generally appreciated these AI-generated rule suggestions. They found them helpful in surfacing overlooked issues and valued the way AI translated group-level insights into actionable rules. For instance, P15, the admin of a semi-formal ex-employee group, described this as  \textit{``the AI does the thinking for me''}, appreciating that the AI not only identified a problem but also suggested a rule and a way to implement it. 

Still, admins emphasized the importance of keeping humans in the loop. Those managing groups with multiple co-admins preferred to vet AI-suggested rules collaboratively, either by discussing with co-admins or posting the suggestion in the group for feedback. 
P16, the admin of a large friends group, explained that she would seek consensus before adopting a rule the AI proposed. 
\textcolor{blue}{Admins like P16 viewed this collaborative vetting as essential because these AI-generated rules do not originate from their own judgment, deliberation with co-admins, or requests from group members. Admins lacked confidence in the AI’s ability to grasp the nuances of their groups and saw participatory review as a way to ensure legitimacy and buy-in.}


In contrast, admins of small, stable groups found the idea of AI-suggested rules redundant. These groups were typically governed by long-standing implicit norms 
and rarely required new formal rules. As P10, the admin of a small friends group, observed:
\begin{quote}
\textit{``I think it [AI] would quickly run out of things to suggest because our group is already well behaved.''}
\end{quote}


Overall, while participants saw potential in AI-assisted rule maintenance—particularly in surfacing trends or overlooked issues—there was limited enthusiasm for automated rule updates. Admins emphasized the importance of human judgment, relational knowledge, and collective input when deciding whether and how to update group \textcolor{blue}{rules}. Until AI systems better support participatory processes and offer contextual cues rather than prescriptive changes, their role in rule maintenance may remain secondary to human-led governance.

\section{Discussion}

Through a combination of speculative design study and co-creation of rules with Meta AI, 
our findings reveal that admins assess the efficiency of AI-assisted rule-making based on how such tools would interfere with the group's context, relational dynamics, and their agency. 
In this section, we critically reflect on these nuances and situate our findings within the broader context of AI-assisted content moderation. 

\subsection{Negotiating Human-AI Boundaries in Online Governance}
While AI assistance can support multiple stages of the rule lifecycle, our findings show that admins do not simply accept or reject AI assistance. Rather, they actively negotiate the boundaries of AI involvement in group governance. These negotiations revolve around four concerns: preserving community privacy, maintaining relational authority, using AI to manage interpersonal tensions, and resisting cultural misalignment. Together, these findings illustrate how AI-assisted moderation is shaped not only by technical capabilities but also by social relationships, power, and cultural context.


\subsubsection{Concerns Around Data Privacy}
Across different WhatsApp groups, admins felt that granting AI access to group chat to create rules was unnecessary and invasive, especially when group metadata could suffice. These concerns echo broader critiques of automated moderation systems, which argue that such systems deepen platform surveillance while limiting user agency and transparency\cite{Gorwa2020, Gillespie2020}. Consistent with the call for visible, accountable moderation systems \cite{Wright2022}, some admins in our study expressed a strong preference for on-device processing, selective data sharing, and the ability to revoke AI access when needed. Rather than viewing data access as a one-time consent process, admins demanded dynamic, revocable, and context-sensitive approach that better preserves the privacy of their communities.  Yet, current AI-assisted moderation tools rarely support this level of control, revealing a critical gap between community expectations and prevailing governance design norms.

\subsubsection{Boundary Work and the Demarcation of Relational Authority}

Across all group types, admins expressed the need to maintain human authority over AI-suggested rules, worrying that AI lacked the ability to interpret sarcasm, multilingual content, and cultural references. By insisting that rule enforcement requires "human touch" and "emotional intelligence", admins categorized these tasks as outside the jurisdiction of AI. We interpret this insistence as a form of active boundary work \cite{gieryn1983boundarywork}, which Gieryn described as the efforts experts undertake to distinguish their professional domain from non-expertise, thereby protecting their autonomy and authority. Drawing on this analogy, we found that WhatsApp admins distinguished between informational tasks, such as summarizing violations and flagging trends, which they were willing to delegate to AI, and relational enforcement, which they viewed as an exclusively human responsibility.


In the context of WhatsApp, this demarcation serves a dual purpose. First, it deters impersonal, context-insensitive AI-suggested rules from damaging social ties. Second, it preserves the admin's status as the group’s indispensable moderator and promotes accountability by limiting overreliance on AI \cite{Jhaver-2019, Gorwa2020, Lai-2022}. Viewed through this lens, AI-suggested rules comprise a form of \textit{boundary objects} \cite{star_griesemer1989boundaryobjects}, which Star and Griesemer described as entities that inhabit multiple social worlds and satisfy the informational requirements of each. Originating in the technical realm of the LLM, they acquire meaning only through the interpretive work of group admins. Thus, effective AI-assisted moderation depends not on replacing human judgment, but on supporting the translation of technical outputs into socially meaningful rules.

\subsubsection{AI as a Social Buffer} 

Although admins sought to preserve authority over rule enforcement, they did not necessarily view AI as a threat. Instead, some saw AI as a useful intermediary that could help manage the interpersonal tensions inherent in moderation. Prior work shows that disagreements over rule enforcement can impose significant  psychological burdens on moderators, often leading them to either quit~\cite{schopke_gonzalez2024moderators} or defer to the preferences of senior moderators~\cite{Cai-2022}. 
In WhatsApp groups, where co-admins can often be family members or peers, the interpersonal stakes of such conflicts are especially high. 
However, rather than quitting or acquiescing to others' preferences,  our participants expressed a desire to use AI to depersonalize conflicts with co-admins. By delegating the \textit{bad cop} role to an AI agent, admins hoped to enforce rules while protecting their personal relationships from the friction of moderation. Admins perceived AI assistance to be a buffer protecting them from relational conflicts, which \citet{schopke_gonzalez2024moderators} identify as a key threat to group's sustainability.

\subsubsection{Accounting for Cultural Misalignment and Algorithmic Hegemony}

Our participants consistently raised concerns about AI failing to grasp the cultural references, relational nuances, and linguistic patterns in group chats. Although our speculative design approach did not allow us to systematically audit specific model failures, the deep-seated skepticism participants expressed points to a broader, well-established challenge in AI-assisted moderation: the normative gaze of Western-centric model.  Research shows that 
such tools often embody a \textit{colonial impulse}, enforcing Western sensibilities as universal standards even when deployed in the Global South~\cite{Shahid2025ThinkOT, Dourish-and-mainwaring, shahid2023decolonizing}. Being predominantly trained on English-centric data, when these models are exported to non-Western contexts, they risk enacting a form of algorithmic hegemony where local communication styles get flagged as deviant simply because they do not align with the Western notions of trust and safety \cite{Sambasivan-reimagining, Shahid2025ThinkOT}. Thus, the widespread adoption of such tools may push users to self-censor their expressions to avoid algorithmic flagging \cite{Gillespie2020}. 

Additionally, current NLP technologies which are primarily designed for English morphology, frequently struggle to process code-mixed content, which is widespread in Indian online communities \cite{Shahid2025ThinkOT}. 
Although our participants interacted with Meta AI only in English and did not share real group chats during the study, they repeatedly raised concerns about how such systems would perform once deployed on actual chats from their WhatsApp groups which often contain code-mixed, multilingual content.  As already outlined in our findings, participants worried that sarcasm and other culturally situated banter, especially when expressed in non-English or mixed-language forms, could be misread as spam or offensive content. If an AI assistant cannot distinguish between benign community banter and actual harassment, it forces admins to perform constant error-correction, increasing their workload rather than reducing it.

\subsection{Design Implications for AI-Supported Group Moderation}
Building on our findings, we argue that AI-supported moderation should augment rather than replace manual governance. Because moderation in WhatsApp groups is deeply relational and often relies on informal social norms, effective AI systems must prioritize contextual understanding, human authority, collaborative control, privacy, and cultural competence. Rather than acting as autonomous moderators, AI systems should function as collaborative partners that support admins in creating, communicating, and enforcing rules while preserving the social fabric of their communities.


\subsubsection{Supporting Contextual Rule Co-Creation, Customization, and Communication} 
Many participants struggled to provide sufficient context for generating appropriate rules and found themselves frustrated by generic AI outputs that required extensive editing. 
Such problems arise because under-specified prompts lead models to overcompensate, burdening users with the task of filtering out irrelevant or verbose content \cite{briakou-2024-verbose}.  Existing 
automated tools such as Reddit’s AutoModerator\cite{reddit2025automoderator} or Facebook’s Admin Assist\cite{facebook2025adminassist} similarly require admins to define rules without explicitly eliciting information about community norms or context.

To support effective rule co-creation, platforms should provide mechanisms for collecting group context through structured interactions. 
These could include brief questionnaires, editable templates, or  conversational interfaces that prompt admins to describe group size, composition, level of formality, common problems, and existing rules. Such context collection should not be a one-time task. Admins should be able to revisit and revise these inputs as group dynamics evolve, ensuring that AI-generated rules remain relevant  over time.

While platforms like Reddit and Discord allow moderators to define rules in editable text formats (e.g., YAML files or keyword lists), these are primarily designed for internal configuration, not member-facing communication. Facebook groups support pinned rules \cite{facebook2025pinrules}, but with minimal formatting and no support for tailoring tones. Since rules ultimately serve as member-facing artifacts, platforms should also make it easy for admins to adapt AI suggestions for communication with group members.  This includes enabling direct editing, providing controls for tone (e.g., formal, friendly, or strict), and supporting formatting options such as bullet points, highlighted text, or emojis. Outputs should also be easily exportable into formats suitable for sharing within groups.

\subsubsection{Preserving Human Authority and Supporting Collaborative Governance}
Consistent with our findings on boundary work, AI governance tools should be designed to empower admins without compromising their agency and control. Rather than automating decisions about rule violations, platforms should give admins the option to selectively delegate tasks to AI, such as surfacing problematic trends, while retaining control over decisions requiring trust, contextual knowledge, or relational judgment. 
Current moderation tools, such as Reddit’s AutoModerator \cite{reddit2025automoderator} acts only on rules explicitly defined by moderators, while Twitch’s AutoMod \cite{twitch2024automod} queues flag messages for moderator approval. Although these systems preserve individual discretion, they offer limited support for collaborative oversight. While multiple moderators may have technical access to rule configurations, there are few built-in affordances for co-editing, discussing, or reviewing moderation decisions collectively. As a result, coordination often happens informally, outside the platform. 

We therefore recommend that AI-supported moderation systems explicitly support collaborative governance. Co-admins should be able to jointly edit AI-generated rules, discuss proposed changes, review moderation decisions collectively, and solicit feedback from community members before adopting new policies. Such mechanisms can help ensure that governance remains transparent and accountable.


\subsubsection{Reducing Interpersonal Burden of Rule Enforcement}
Our findings show that moderation is not merely a technical process but also a relational one. Admins in small, informal groups often preferred direct communication with rule violators to preserve social relationships. In contrast, admins of semi-formal groups were more open to AI-mediated intervention following rule violation to reduce personal backlash and interpersonal tension. 
Although existing tools such as Reddit’s AutoModerator\cite{reddit2025automoderator} or Discord’s AutoMod \cite{discord2024automod} can remove content automatically, they do so without any relational consideration. Additionally, none of these tools provide editable templates or writing support for admins to send personalized warning to violators or the whole group.

To support relational moderation, platforms should provide customizable templates and AI-assisted writing support that help admins communicate warnings, explain rules, or address problematic behavior. These systems should preserve cultural nuances and interpersonal dynamics rather than imposing generic or impersonal language. Supporting relational communication may allow AI to function as a buffer that reduces interpersonal friction without removing humans from the moderation process.

\subsubsection{Designing for Privacy and Cultural Competence}
Our findings suggest that privacy and cultural alignment are essential prerequisites for trustworthy AI-supported moderation. In WhatsApp, end-to-end encryption limits server-side access to user-generated content \cite{whatsapp2025e2e}, making unrestricted access to group conversations both technically challenging and socially undesirable. We therefore recommend that AI systems rely primarily on local processing or structured metadata supplied by admins rather than full chat histories 
which may erode trust, especially in sensitive settings. More broadly, on platforms like Reddit and Discord, where content is publicly accessible, we recommend that AI governance tools incorporate granular and revocable controls for content while clearly communication with group members. 

Designers must also account for cultural and linguistic diversity. Western-centric AI models frequently misinterpret non-Western communication styles, flagging benign content as toxic~\cite{Shahid2025ThinkOT} and may suggest rules irrelevant to the local context. Therefore, we recommend that AI-assisted moderation tools explicitly support local language customization. Rather than solely relying on pre-trained Western standards, platforms should allow admins to provide few-shot examples of acceptable local speech, including code-mixed slang, playful banter, or community-specific terms of endearment, to guide the rule drafting and enforcement process. By allowing admins to define the boundary between toxicity and camaraderie within their local context, designers can avoid silencing legitimate speech and ensure that AI respects the community values rather than enforcing Western standards.

\subsection{Limitations and Future Work}

Our study has several limitations that open avenues for future research. First, we used a design probe to illustrate AI-assisted moderation features. Thus, our findings reflect perceived value of such interventions rather than actual outcome, which is an inherent limitation of design probes. This speculative approach was necessary given the ethical and practical constraints of deploying a live AI assistant in encrypted WhatsApp groups. 

Second, participants co-created rules using the Meta AI chatbot (Llama 3.2) integrated within WhatsApp. 
While this allowed us to examine admins’ interactions with an LLM-chat bot when co-creating rules, 
we cannot distinguish whether observed frictions, such as verbosity or rigid formatting, are inherent limitations of generative AI or specific to Meta AI chatbot. Future studies should explore how different models and prompting strategies may influence admin experiences when co-creating rules.

Third, our sample skewed toward urban, college-educated WhatsApp users in India due to recruiting limitations related to conducting remote research. Since WhatsApp is widely used across diverse populations, future work should recruit a broader demographic, including rural, non-English speaking, and lower-literate users, to get a full picture of how AI moderation tools are perceived across diverse social contexts.

Fourth, many groups in our study were collaboratively governed by multiple admins. However, we only recruited one admin from each group due to logistical complexity of coordinating multiple admins from the same WhatsApp group. 
Additionally, several participants expressed that they would not be comfortable discussing group dynamics in the presence of fellow admins. This limited our ability to systematically examine how power dynamics among admins, conflicting moderation styles, or disagreements over AI-generated rules might shape decision-making.


Finally, future research could actually develop AI-assisted rule creation tool to assess their outcome. Given end-to-end encryption and restrictions around deploying external tools within  WhatsApp, future systems might explore off-platform tools (e.g., a website for rule drafting and sharing) that decouple governance from platform's own infrastructure. Such tools could employ a data donation model building on tools like \textit{WhatsApp Explorer} \cite{GarimellaChauchard2025WhatsAppExplorer}, where admins can voluntarily share existing group rules or group data to get rule suggestions that are specific to their groups. 
Longitudinal deployments could help evaluate how admins’ trust, preferences, and usage evolve over time, with shifting group dynamics and in the context of multi-admin collaboration. 

\section{Conclusion}
Our study highlights how WhatsApp group admins across diverse groups perceive, adapt, and resist AI assistance in group rule creation and moderation. While admins of large, formal  and semi-formal groups saw AI tools as useful for drafting rules, summarizing violations, and handling scale, admins of smaller, close-knit groups often found such interventions unnecessary or relationally disruptive. Across the spectrum, participants asserted the importance of preserving human authority, relational sensitivity, and contextual judgment, especially in emotionally charged or ambiguous scenarios. These findings challenge assumptions that automation universally eases moderation, showing instead that AI-assisted moderation tools must flexibly adapt to different governance styles, group dynamics, and expectations of control. 

We offer design recommendations focused on providing structured mechanisms to collect relevant group context from admins to generate appropriate rules, enabling flexible customization of tone, length, and formatting as per the group culture, supporting privacy-conscious data sharing and revocable consent, and building features that allow admins to collaboratively review and approve AI outputs with co-admins and group members before enforcement. \textcolor{blue}{Ultimately, we advocate for a flexible design where AI tools remain optional, allowing small groups to function without unnecessary automation.} As platforms begin exploring AI-driven tools for moderation in end-to-end encrypted environments, our work underscores the need to design AI-assisted tools that support admins' relational practices, accommodate diverse group \textcolor{blue}{rules}, and preserve human judgment, especially in private, small-group settings.

\vspace{1em}
\noindent\textbf{\large Acknowledgments} \\[0.5em]
We sincerely thank the participants for sharing their thoughtful reflections on how AI can support the creation, enforcement, and maintenance of group rules in WhatsApp communities. This work was supported in part by Infosys and NSF Grant \#2542732.

\bibliographystyle{ACM-Reference-Format}  
\bibliography{main} 

\newpage
\appendix
\raggedbottom

\section{Codebooks}
\noindent
\textcolor{blue}{The following tables present the final codebooks used in our analysis. The Phase 1 codebook captures admins’ interactions with Meta AI during rule co-creation, while the Phase 2 codebook captures participants’ reflections on AI-assisted moderation elicited through interviews and design probes. For each codebook, we provide code definitions and their corresponding sub-themes.}
\begin{table}[h]
\centering
\small
\setlength{\tabcolsep}{5pt}
\renewcommand{\arraystretch}{2.18}
\caption{Phase 1 Meta AI interaction-log codebook: admin--AI rule co-creation behaviors (Subthemes 1--7).}
\label{tab:phase1_interaction_codebook_b1}

\begin{tabular}{p{0.22\textwidth} p{0.28\textwidth} p{0.42\textwidth}}
\toprule
\textbf{Subtheme} & \textbf{Codes (examples)} & \textbf{Definition} \\
\midrule

Minimal context / underspecification &
No initial context; AI asks for context; generic rules without context &
Admin begins with vague prompts or missing group details, resulting in generic outputs and/or AI requests for clarification. \\ \addlinespace

Progressive context building &
Group metadata; group setting; existing rules; will share more details &
Admin incrementally supplies group information (purpose, norms, constraints) to steer outputs toward fit-for-group rules. \\ \addlinespace

Expectation of group-specific rule discovery &
Wants new group-specific rules; AI expands rule coverage &
Admin expects AI to surface overlooked or novel rules beyond generic defaults; AI is treated as a source of additional coverage. \\ \addlinespace

Incorporating existing policies into AI output &
Adds existing policies; provides existing rules as context &
Admin uses AI output as a draft but ensures alignment with pre-existing group policies and incident-driven norms by adding missing rules. \\ \addlinespace

Tone \& language alignment &
Language/tone alignment; appreciates language &
Admin requests changes to make wording socially appropriate (formal/strict/casual), accessible, and aligned with the group’s norms. \\ \addlinespace

Conciseness / reducing verbosity &
Asks for shorter rules; dislikes long rules &
Admin edits long lists into fewer, shorter, and more scannable rules, prioritizing legibility and adoption. \\ \addlinespace

Relevance filtering / removing generic rules &
Removes irrelevant rules; generic broad rules &
Admin identifies and deletes rules perceived as generic, redundant, or misaligned with the group’s purpose and practices. \\

\bottomrule
\end{tabular}
\end{table}

\begin{table}[H]
\centering
\small
\setlength{\tabcolsep}{5pt}
\renewcommand{\arraystretch}{1.18}
\caption{Phase 1 Meta AI interaction-log  codebook: admin--AI rule co-creation behaviors (Subthemes 8--14).}
\label{tab:phase1_interaction_codebook_b2}
\begin{tabular}{p{0.22\textwidth} p{0.28\textwidth} p{0.42\textwidth}}
\toprule
\textbf{Subtheme} & \textbf{Codes (examples)} & \textbf{Definition} \\
\midrule

Iterative rule editing (targeted modifications) &
Modifies a rule; incorporates feedback; AI invites edits/additions &
Back-and-forth revisions where admin requests specific changes and checks whether AI correctly integrates feedback across turns. \\ \addlinespace

Formatting for shareability (post-ready text) &
Post-ready version; numbering; spacing; highlighting &
Admin requests layout fixes that produce a clean copy-pasteable message suitable for posting in the group (structure, emphasis, spacing). \\ \addlinespace

Aesthetic / visual presentation &
Emojis; post-ready image of rules &
Admin requests stylistic treatments (emojis) or alternative renderings (image format) to match the group “vibe” or platform constraints. \\ \addlinespace

Rule communication scaffolding (with skepticism) &
AI suggests communication without asking; admin skeptical &
AI proactively suggests communication strategies (e.g., reminders framing), but admin expresses hesitation about appropriateness or social risk. \\ \addlinespace

Seeking enforcement strategies beyond drafting &
Asks AI for enforcement strategies &
Admin extends beyond drafting rules to ask how rules should be enforced or responded to after violations. \\ \addlinespace

Low-effort drafting benefits &
Low-effort input; saving time writing rules &
Admin explicitly values AI as labor-saving support for drafting and structuring rules quickly. \\ \addlinespace

Interaction friction / breakdowns &
Frustrated by AI; context+edit difficult; thinks AI fulfilled request &
Admin experiences breakdowns in the interaction (difficulty steering edits, mismatched outputs) and/or reaches a stopping point when the goal feels complete. \\

\bottomrule
\end{tabular}
\end{table}

\begin{table}[!t]
\centering
\small
\setlength{\tabcolsep}{5pt}
\renewcommand{\arraystretch}{1.18}
\caption{Phase 2 codebook for interaction logs (Subthemes 1--7).}
\label{tab:phase1_codebook_a1}
\begin{tabularx}{\columnwidth}{p{2.5cm} p{3.0cm} Y}
\toprule
\textbf{Subtheme} &
\textbf{Merged codes included (examples)} &
\textbf{Definition} \\
\midrule

Conditional usefulness by group type &
AI moderation useful for large groups vs unnecessary for small groups; AI rule suggestions useful/necessary vs unnecessary; reports valuable for large groups; moderation summaries useful vs not useful &
Participants evaluated AI support as situational: high value for high-volume or low-relational groups (formal/semi-formal), and low value in small, close-knit groups where manual governance is already sufficient. \\

Language \& localization constraints &
Concern: local language moderation; concern: inability to understand nuance &
Participants questioned AI reliability due to multilingual content, mixed scripts, slang/sarcasm, and culturally embedded references that may be misinterpreted or incorrectly flagged. \\

Privacy \& data locality preferences &
Concern: data privacy; preference for local/on-device AI; lack of data usage terms; member opt-out and consent &
Privacy governance was framed as foundational: participants wanted transparent data-use terms, local processing where possible, and dynamic, revocable consent. \\

Selective data access boundaries &
Complete data access; cannot share member-generated content; selected access: metadata/links/messages; selected access: images+memes (fandom) &
Participants negotiated granular boundaries around data-sharing, distinguishing low-risk inputs (metadata/links) from sensitive user-generated content (messages/images), often preferring minimal access. \\

Human authority as primary decision-maker &
Need for human agency as primary authority; reports require context; human verification of AI-flagged content &
Participants consistently framed AI as assistive: admins must retain control over final moderation decisions, enforcement, and accountability. \\

Relational moderation / ``human touch'' &
Concern: lack of human touch; AI intervention unnecessary when admin personally interacts; no AI moderation in close groups &
In close-knit groups, moderation is relational labor requiring emotional intelligence and social hierarchy-awareness; AI interventions were perceived as impersonal, disruptive, or legitimacy-threatening. \\

Editing \& customization as core interaction &
Admin wants to make manual edits; minor edits manual/major edits AI; need customized AI moderation &
Participants treated AI outputs as drafts requiring iterative refinement and customization to match group rules, admin style, and intended tone. \\

\bottomrule
\end{tabularx}
\end{table}

\begin{table}[h]
\centering
\small
\setlength{\tabcolsep}{5pt}
\renewcommand{\arraystretch}{1.18}
\caption{Phase 2 codebook for interaction logs (Subthemes 8--14).}
\label{tab:phase1_codebook_a2}
\begin{tabular}{p{0.22\textwidth} p{0.28\textwidth} p{0.42\textwidth}}
\toprule
\textbf{Subtheme} &
\textbf{Merged codes included (examples)} &
\textbf{Definition} \\
\midrule

AI as rule-writing assistant (pragmatic benefits) &
AI support for lack of English proficiency; saving time; AI rule suggestions useful and necessary &
Participants valued AI for drafting structured rules quickly and improving wording/clarity, reducing the cognitive and linguistic burden of authoring and formatting rules. \\

Legitimacy via co-admin/member involvement &
Verification with co-admins; collaboration with members; disagreements among admins/members; multiple co-admins &
Participants emphasized collective vetting for legitimacy and buy-in, especially for AI-suggested rules, through co-admin coordination or member input. \\

Existing moderation practices: monitoring \& maintenance &
Quick daily review; passive moderation when notified; active moderation when new members join; deleting irrelevant/offensive content; keeping conversations relevant; preventing DMs &
Participants described routine governance as a mix of proactive monitoring and reactive interventions shaped by group volume, admin workload, and evolving group composition. \\

Rule communication channels \& implicit rules &
Posting in group; posting in description; assuming implicit understanding &
Participants communicated rules via lightweight infrastructures (posts/description) or relied on implicit shared rules, especially in stable or close-knit groups. \\

Rule creation triggers \& content harms &
Conflict among members; hate speech; spam posting; members don’t obey rules &
Rules were often formalized reactively in response to incidents involving spam, hate speech, repeated conflict, or recurring boundary violations. \\

Violation response strategies \& escalation ladder &
Admin DMs offender / calls / in-person; AI draft DM; AI DM with verification; group reminders (or avoided); multiple chances; temporary bans; removal &
Participants described graduated enforcement: private vs public interventions, escalation severity, and relational consequences. AI was primarily acceptable as a drafting aid or buffer with oversight. \\

Comparison to existing automation ecosystems &
Reddit AutoMod perceived superior; Discord bots perceived superior &
Participants compared WhatsApp’s limited tooling with other platforms’ mature moderation infrastructures, shaping expectations for configurability and automation. \\

\bottomrule
\end{tabular}
\end{table}

\clearpage
\end{document}